\documentclass[aps,amsmath,preprint,showpacs,preprintnumbers]{revtex4}
\usepackage{graphicx}
\usepackage{dcolumn}
\usepackage{multirow}
\usepackage{slashbox}
\usepackage{subfigure}
\usepackage{bm}

\begin{document}


\title{A Microscpoic Model for the Neutron Dynamic Structure Factor of Solid Methane in phase II}

\author{Yunchang Shin}%
 \email{yunshin@indiana.edu}
 \homepage{http://www.iucf.indiana.edu}
\author{W. Mike Snow}
\author{Chen-yu Liu}
\author{Christopher M. Lavelle}
\author{David V. Baxter}
\affiliation{Department of Physics, Indiana University/IUCF, 2401
Milo B. Sampson Lane, Bloomington, IN 47408, USA}

\date{\today}

\begin{abstract}
We have constructed an approximate  microscopic model for the neutron dynamic structure factor of solid methane in phase II. We expect our model to apply for neutron energies below 1\textit{eV} at pressures near 1 bar and temperatures below 20K, where solid methane exists in a phase, called phase II, in which the tetrahedral hydrogens exist in a mixed phase of free rotors and hindered rotors. Certain matrix elements needed for the dynamic structure factor $S(Q, \omega)$ are adapted from a calculation by Ozaki \textit{et~al.}\cite{ozaki:3442}\cite{ozaki2}. The model treats the effects of molecular translations,  intramolecular vibrations and the free and hindered rotational degrees of freedom as uncoupled ones.  Total scattering cross sections calculated from the model agree  with the cross section measurements of Grieger\cite{grieger:3161} and Whittemore\cite{Wittemore} for the incident neutron energies of  0.5 \textit{meV} $\sim$ 1\textit{eV}. The effective density of states extracted from the model was compared with the Harker $\&$ Brugger frequency spectrum extracted from neutron scattering measurements conducted at 22K\cite{HARKER1967}.
\end{abstract}

\pacs{28.20.Cz;28.20.Gd;.29.25.Dz;78.70.Nx}
\maketitle

\section{\label{sec:1 }Introduction}


The theory of phase space compression of neutron ensembles, usually called moderation in neutron physics, has been well-developed for nuclear engineering and related applications in the energy regime which neutrons possess upon liberation from nuclei (\textit{MeV}) to the kinetic energies of atoms in matter at room temperature (~25 \textit{meV}). Since these energies are large compared to the kinetic energy and the binding energy of the atoms in the moderating medium, the total cross section is given to a good approximation by the incoherent sum of the scattering cross sections from the individual atoms in the medium. In this limit, the theory of neutron moderation needs only to apply energy and momentum conservation to a sequence of collisions of neutrons of mass $m$ and initial energy $E_i$ with target nuclei of mass $M$ and zero kinetic energy. Such an analysis forms the core of neutron moderation theory in nuclear reactors. The large number of collisions typically needed to reduce the energy of a neutron by several orders of magnitude leads inevitably to neutron emission sources which are incoherent.

For neutron energies at or below the energies of motion of the moderating medium, however, the situation is qualitatively different. In this case, the neutron wavelength is larger than the separation between atoms in the medium, and in this case the scattering exhibits strong interference effects which depend on the details of the structure and modes of motion of the medium. Furthermore, the  inelastic fraction of the total cross section which is, of course, the essential component for phase space compression, decreases as the neutron energy drops below the binding energy and the $~meV$ excitation energies of collective modes in the medium. Interaction of the neutrons with external fields are of little help in phase space compression: the gravitational interaction is conservative, the weak interaction is too weak, and the small magnetic moment of the neutron leads to inelastic effects from neutron spin flips in external magnetic fields in the $\mu eV$ range, which is too small to be practical as a moderation technique for neutrons with kinetic energies in the \textit{meV} range. Therefore, in practice, the only efficient way at present to increase neutron phase space density is through collisions in a medium.

The amount of phase space compression per collision for slow neutrons in a moderating medium,  therefore, tends to decrease as the energy decreases. Although one can in principle increase the phase space density in a neutron moderator indefinitely in the absence of absorption by cooling the medium to $T=0$ and allowing the neutrons to undergo an arbitrary number of collisions, in practice the finite neutron absorption cross section of all media other than $^{4}He$ places an upper bound on the number of collisions that can be tolerated, and then once again the amount of phase space compression per collision depends on the microscopic properties of the medium even at $T=0$. In addition the inelastic modes available in a condensed medium tend to freeze as $T \to 0$, thereby reducing the efficiency of the moderating medium even further.

The rate of phase space compression per collision is proportional to the neutron dynamic structure factor $S(Q,\omega)$ in the medium, where $\omega=E_{i}-E_{f}$, $\vec{Q}=\vec{k_{i}}-\vec{k_{f}}$ and $E_{f}$, $E_{i}$, $\vec{k_{f}}$, $\vec{k_{i}}$ are the final and initial neutron energies and momenta. In the theory of neutron phase space compression in the cold neutron regime, we are especially interested in the behavior of $S(Q,\omega)$ as $\omega \rightarrow 0$. This is because the width of the distribution of neutrons in phase space that we are interested in cooling is typically large compared to the range in $(Q,\omega)$ space over which $S(Q,\omega)$ is large. When the latter range is defined by the dispersion relation of a well-defined elementary excitation of the medium such as a phonon, magnon, libron, etc., $S(Q,\omega)$ is large only over a narrow ridge in $(Q,\omega)$ space. If the neutrons of interest for cooling also possessed a phase space distribution of comparable width, the obvious optimal solution to neutron cooling would be to match the elementary excitation spectrum to the neutron phase space to be cooled and reduce the neutron energy to zero by the creation of a single elementary excitation, with the refrigerator that maintains the moderating medium at $T=0$ providing the dissipation required for phase space compression. This strategy is used in so-called superthermal neutron moderators optimized for the creation of ultracold neutrons. For cooling a broader phase space distribution of neutrons to energies in the $100 \mu eV \sim meV$ range, however, multiple collisions are essential, and $S(Q,\omega)$ must, therefore, possess some strength near $\omega \rightarrow 0$ so that most neutrons in the distribution have energy losses $\omega_{i}$ in some sequence of collisions $S(Q, \omega_{i})$ that can allow them to approach $E_{f} \rightarrow 0$.

These considerations guide the search for improved neutron moderating media into
some obvious directions. First of all, it is clear that normal $3D$ phonon excitation becomes inefficient at low energies, since the density of states and therefore $S(Q,\omega)$ vanishes as  $\omega \to 0$. The abnormally
large scattering cross section of hydrogen, typically more an order of magnitude larger than for other nuclei, makes hydrogenous materials the obvious choice if the
number of collisions for phase space cooling is not limited by the $0.3$ barn neutron absorption cross section of hydrogen at $25 meV$ and its $1/\sqrt{E}$ increase at lower energies.

Solid methane is a special case of particular importance for cold neutron moderation theory for a number of reasons. First of all, solid $CH_4$ possesses a
high number density of hydrogen atoms. Second, the high symmetry of the $CH_4$ molecule is ultimately responsible for one of its most interesting properties: namely, that a significant fraction of the tetrahedron of $H$ atoms in $CH_4$ are free to rotate even close to $T=0$ in the solid phase. This property is shared by few other molecular solids  such as solid hydrogen and $\gamma$-picolene \cite{PressBook}. The resulting inelastic free rotor modes are responsible for the observation, confirmed empirically through measurements of neutron spectral intensity at spallation neutron sources, that solid methane is the brightest cold neutron moderating medium\cite{Conrad}. Third, the motions of the rotational, translational, and vibrational modes of $CH_4$ are sufficiently decoupled and the molecule is simple enough in structure that one can foresee the possibility of performing a theoretical calculation of the dynamic structure factor $S(Q,\omega)$ of reasonable accuracy from first principles\cite{grieger:3161}. Such an analytical treatment in the theory of neutron scattering can be more easily used to gain physical insight into the cold neutron moderation process. In particular, one can analyze quantitatively some suggestions that have been made in the past to modify the $CH_4$ medium to increase its cold neutron brightness. For example, in the $T \leq 20$ phase of solid methane, phase II,  only 1/4 of the tetrahedral are free to rotate, and 3/4 of the sites undergo libration and tunneling motions. Many have speculated that one could improve the cold neutron brightness from a methane moderator if one could modify the $CH_4$ in some way so as to free up the rest of the modes.
 However, the librational and tunneling motions also possess inelastic components, and it is not obvious which of the two modes presenting in phase II solid $CH_4$ is, in fact, more effective in neutron cooling, although the commonly-accepted folklore is that the rotational mode is the essential component.
Fortunately the librational and tunneling energy spectra for phase II solid $CH_4$ have been measured\cite{huller:4495}\cite{huller}, and so the theory need not attempt to perform the delicate calculation to predict these effects from first principles but rather can simply take them from experiment. Yet another theoretical simplification comes from the
identity of the 4 hydrogen atoms, which correlates the form of the rotational and spin wave functions through the overall symmetry they must possess under identical particle interchange.

Another consideration which makes this work timely is the development of new types of bright cold neutron sources. At high-power neutron spallation neutron sources,
the radiation damage in solid methane from fast neutrons and gammas is so severe that a sequence of events is initiated which leads in short order to the catastrophic warm-up of the moderator. The Low Energy Neutron Source (LENS) at Indiana University Cyclotron Facility (IUCF) is a long pulsed accelerator based neutron source based on $(p,n)$ reaction in the $Be$ target. The low heat load on the LENS moderator due to the greatly reduced number and energy of fast neutrons and capture gammas in this type of source make it possible to operate our solid methane moderator, and also other ¡°delicate¡± moderating media which would be destroyed if operated in a MW-power spallation source environment, at temperatures well below 20K with infrequent warm-ups to release stored energy. In addition to its use as a cold neutron moderator at LENS, solid methane is planned to be used as a premoderator for an ultracold neutron (UCN) source at the NC State PULSAR reactor.

In this paper, we present an approximate model for $S(Q,\omega)$ for phase II solid $CH_4$. The rest of the paper is organized as follows. Section II discusses the structure of solid methane in phase II. Section III discusses a neutron scattering model of solid methane. The results including neutron cross section and spectral intensity are on section IV. We conclude and suggest further work in section V.

\section{\label{sec:2} Solid Methane in phase II}

\begin{figure}[htbp]
\begin{center}
\includegraphics[width=10cm]{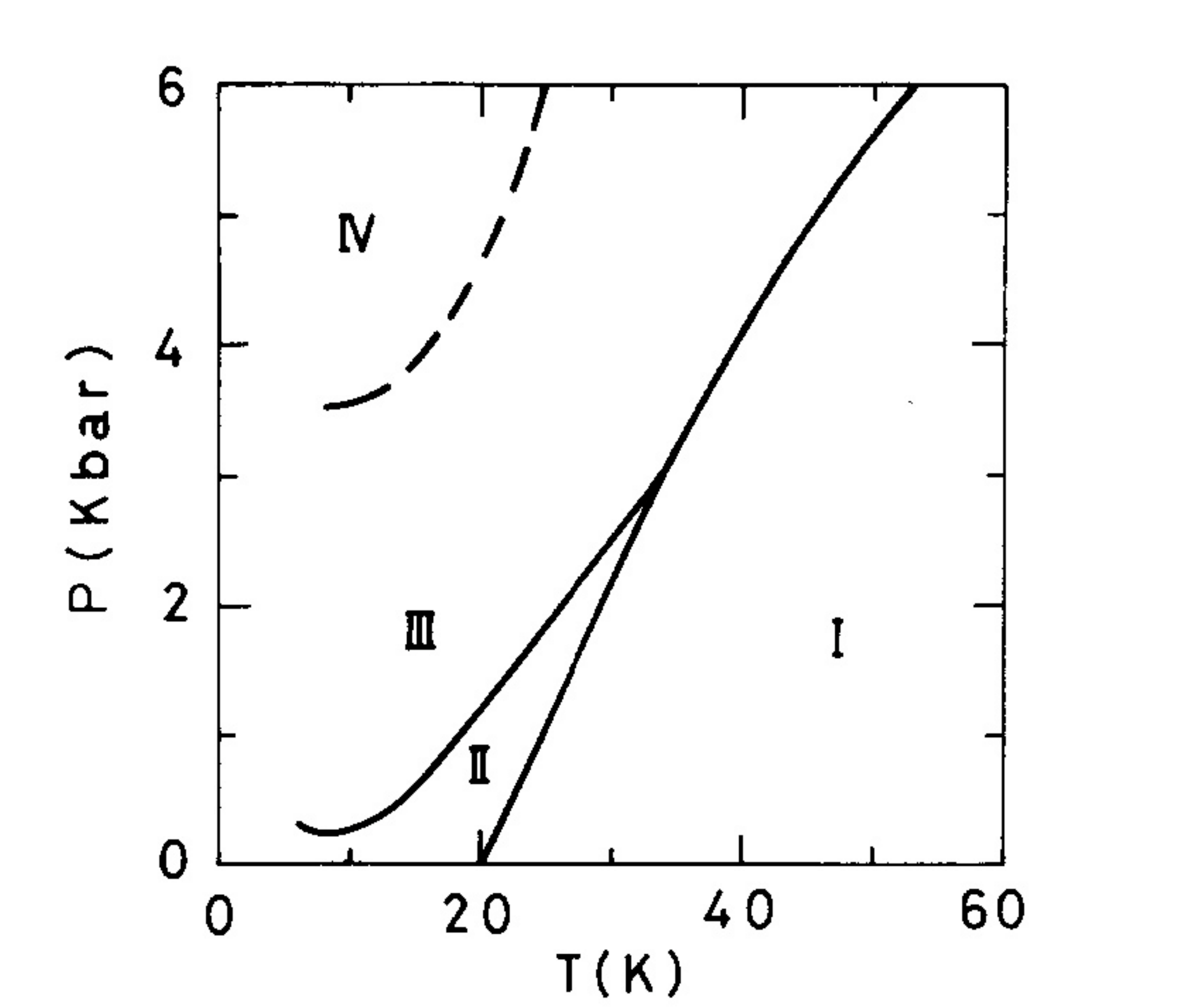}
\caption{The phase diagram of solid methane \cite{yamamoto:2701}}
\label{fig:3.1}
\end{center}
\end{figure}

Solid methane at low temperatures has  two distinct solid phases. The detailed phase diagram is shown in Fig. \ref{fig:3.1}. At equilibrium vapor pressure, phase I appears from the triple point temperature (90.6K) to the transition temperature (20.4 K) into phase II.  In phase I, methane has a face-centered cubic (\textbf{fcc}) structure with four orientationally disordered molecules per unit cell. This rotational motion of tetrahedral methane molecules remains slightly retarded even near the melting point of the crystal\cite{Manzhelii}.

The phase change is correlated with a structural  shift from
the point symmetric group ${O}_h$ in phase I to the
symmetrical group ${D}_{2d}$ and ${O}_h$ in phase II.
In phase II, below 20.4K, the center of mass of methane molecules
is located on the site of the face-centered cubic lattice. However, a
structural transition to a partial orientational order occurs.
This ordering of molecular orientation leads methane to possess  two
different sublattices, ${D}_{2d}$ and ${O}_h$ in phase II
\cite{Manzhelii}.

\begin{figure}[htbp]
\begin{center}
\includegraphics[width=8cm]{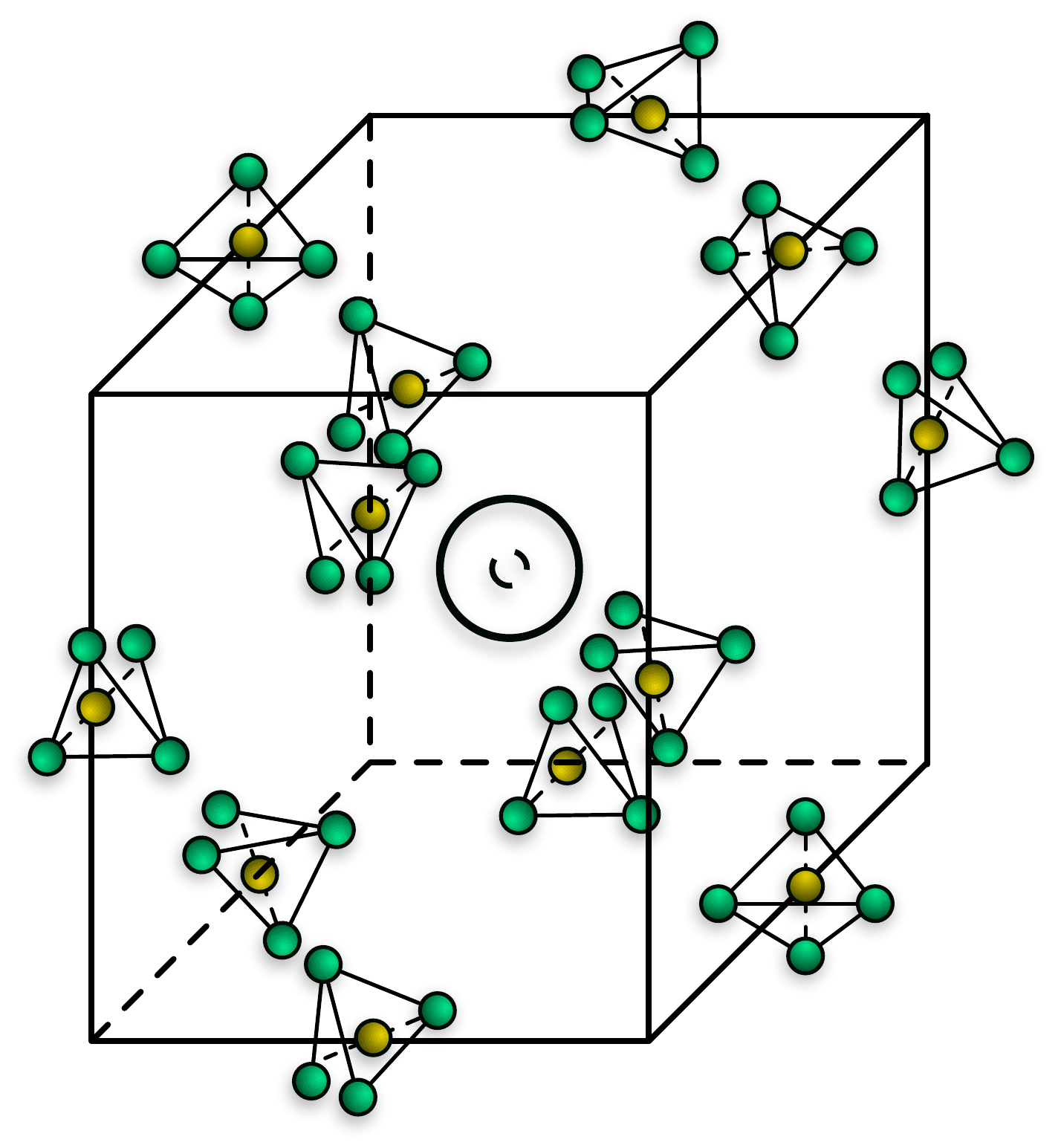}
\caption{The structure of solid methane in phase II\cite{grieger:3161}. The circle one represents almost free rotation of the molecule. The tetrahedra denotes the orientationally ordered molecules in the structure.}
\label{fig:3.2}
\end{center}
\end{figure}

Fig. \ref{fig:3.2} shows the partially ordered structure of phase II. It is called an eight-sublattice antiferorotational structure. The unit cell of phase II contains 32 molecules. Three-fourths
of the molecules are hindered rotors which librate in the  minima
of a strong orientational potential of ${D}_{2d}$ symmetry. the other  1/4 of
the molecules are free rotors.

This structure was predicted by James and Keenan on the basis
of electrostatic octopole-octopole interaction\cite{james:12}.
The
structure of phase II was experimentally determined by W. Press\cite{press:2597}
through  neutron diffraction on  ${CD}_4$. In the solid
${CD}_4$, phase II is stable in the temperature range
22K$<$T$<$27K. Although the direct structure of ${CH}_{4}$
was not possible to determine because of the high incoherent cross
section of protons, they assumed that the experimentally measured
symmetry of ${ CD}_4$ implies  the same orientational
structure of ${ CH}_4$.

\subsection{\label{sec:2-1} Rotational Excitation of Free Rotor: $O_{h}$ symmetry}
In phase II, ${CH}_{4}$ is in a partially ordered state with respect to the molecular orientations. The eight sublattice structure consists of six ordered sites with point symmetry $\bar{4}2m$ and two disordered sites with point symmetry $432$. The disorder  is  caused by the partial or complete loss of  the three dimensional periodicity of physical properties. In a molecular solid, orientational disorder can be produced if a part of the anisotropically interacting molecules is replaced by a isotropic unit\cite{PRAGER:1988lr}. Meanwhile, the ordering is caused by the octopole-octopole interaction. The site of $432$ symmetry feels only a weak crystalline field and the methane molecules are therefore almost free quantum rotors. The free rotor is affected more by disorder because the relaxation of orientations around a defect removes the compensation of octopole-octopole interaction.

\begin{figure}[htbp]
\begin{center}
\includegraphics[width=8cm]{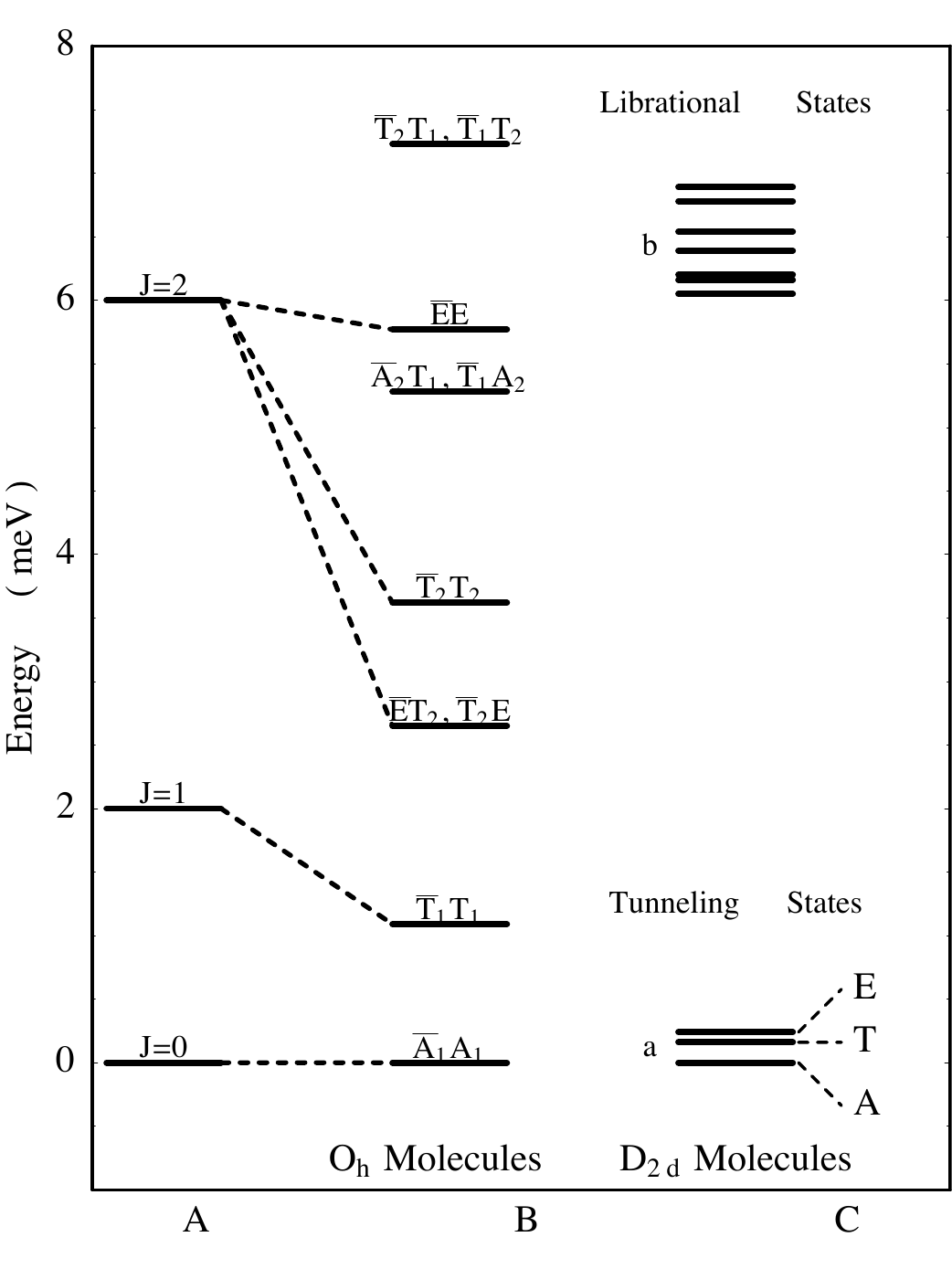}
\caption{The energy levels\cite{2005CoPhC.171..107C} of two different spin states of $\textrm{CH}_{4}$ molecules: (a) $O_{h}$ is orientationally disordered molecules (b) $D_{2d}$ is orientationally ordered molecules}
\label{fig:3.3}
\end{center}
\end{figure}

The Hamiltonian of the Extended James-Keenan (EJK) model for molecule $i$ in
a crystal field $V_{c}$,  from the Eq. 2.15 in Yamamoto.$\it{et~al.}$\cite{ yamamoto:2701}, is
\begin{equation}
H_{i}= K_{i}+f_{c}\sum_{l\geq 4}V_{c}(\omega_{i}^{E})=K_{i}+f_{c}B(\beta_{4}(\omega_{i}^{E})+\beta_{6}(\omega_{i}^{E})).
\end{equation}
The numerical values of $\beta_{4}$ and $\beta_{6}$ are listed in
Table. I of Yamamoto.$\it{et~al.}$\cite{yamamoto:2701}. $K_{i}$ is the operator of the
Rotational kinetic energy. The  parameter $f_{c}=1.25$ was
 determined by  fitting the strength of the crystalline field
\cite{yamamoto:2701}. The rotational wave function is given by
the standard Wigner functions $\mathfrak{D}_{k,m}^{l}(\omega)$. The
rotational states of a tetrahedral rotor with potential parameters
determined for $CH_{4}$ are shown in column B of Fig. \ref{fig:3.3}.
The energy levels
are classified through the  symmetry $\bar{\Gamma}\Gamma $ of
the corresponding wave function. The label $\bar{\Gamma}\Gamma $
denotes for the 25 irreducible representation of the direct
product group $\bar{\textbf{O}}\times\textbf{O}$ which describes
the symmetry with reference to the crystal-fixed frame ($\Gamma$)
and with reference to the molecular fixed frame ($\bar{\Gamma}$).
The total wave function of a ${CH}_{4}$ molecule is the product of a rotational wave function and a spin wave function due to the Pauli permutation principle, in which the total wave function must be
anti-symmetric with a permutation of two protons. Therefore, each
state is related to a representation $A$, $T$ and $E$ corresponding to
the total nuclear spin $I=2$ (ortho), $I=1$(meta) and $I=0$(para),
respectively. The state $\bar{E} E $ cannot be reached by neutron interaction from the
ground state $\bar{A} A $ for the reason of nuclear spin conservation
\cite{PressBook}.

The more simple description with integer concerning the energy level
which begins with $J=0$ for the ground state has been notated by
Ozaki $\textit{et~at.}$\cite{ozaki:3442}. As may be seen in column A and B of Fig. \ref{fig:3.3}, the $J=1$
level is lowered from the free rotor value by $25\%$  from  crystalline fields. There is some ambiguity in $J=2$ state
concerning the energy level splitting into additional states. The energy transitions between rotational energy levels are 1.09 $meV$ in $J=0 \leftrightarrow J=1$ transition, 1.56 $meV$ in $J=1 \leftrightarrow J=2$ transition and 2.65 $meV$ in $J=0 \leftrightarrow J=2$ transition.

\subsection{\label{sec:2-2} The rotational tunneling in molecular field: $D_{2d}$ symmetry}

The molecule on the orientationally ordered lattice point of
${CH}_{4}$ librates around  the
minima of the orientational potential. There is also the additional
possibility to tunnel between each of the indistinguishable
equilibria in two or three dimensional rotational motions of
${CH}_{4}$ molecule. The rotational tunneling describes the
phenomenon of the librational state of a molecule whose rotating
atoms are indistinguishable\cite{prager2}. The splitting between
the states is called ``\textit{Rotational Tunneling}".

The calculation of eigenstates for using free rotational wave
functions was performed  in the framework of EJK model by Yamamoto $\textit{et~al.}$\cite{yamamoto:2701}. These procedures required a diagonalization of an infinite dimensional  Hamiltonian
matrix. Although the procedure converged quickly because of the weak
orientational potential, the wave functions of these tunneling levels are
only slightly different from those of the completely free rotor.

An  alternative approach was developed by H\"{u}ller and Kroll
using a ``pocket state" to treat rotational and tunneling excitation
in molecular crystal\cite{huller:4495}\cite{huller}. A systematic
representation is shown in column C of Fig. \ref{fig:3.3}. A rotational
potential with $n$ minima on two or three dimensional unit sphere was considered .
These minima have a symmetric relation to each other. A set of states
$|i>$ centered at each minimum is taken as the basis set. $|i>$ is the
state of molecule called ``pocket state'' which can find itself in
this basis set. Applying operator $R_{n}$ from the symmetry group
T of the $CH_{4}$ molecule, one obtains the states $|n>=R_{n}|i>$.
Here $R_{n}$ denotes  one of the 12 symmetry elements of the
tetrahedral group T (T=23: subgroup of proper rotation of point
group $\overline{4}3m$). H\"{u}ller explicitly gave the
transformation which diagonalizes the $12\time 12$ Hamiltonian
matrix $<n'|H|n>$ of the  system. For tetrahedral  symmetry, there
is just one $120^{\circ}$ overlapping matrix element
$h=h_{1}=h_{2}=h_{3}=h_{4}$, as well as $180^{\circ}$ overlapping
matrix element $H=H_{1}=H_{2}= H_{3}=H_{4}$. Diagonalization yields
states with $A$,$T$ and $E$ symmetry with the following
eigenvalues
\begin{eqnarray}
E_{A}&=&D+3H+8h~~~(\textrm{singlet})\\\nonumber
E_{T}&=&D-H~~~~~~~~~~~(\textrm{three triplets})\\\nonumber
E_{E}&=&D+3H-4h ~~~(\textrm{doublet}),
\end{eqnarray}
where $D=<n'|H|n>$ is the diagonal element of the
Hamiltonian matrix. In a strong orientational field (in case
$H<<h$), the tunneling of double axes can be neglected\cite{PressBook}.

The oscillation frequency, called the tunneling frequency, depends on
the overlap of pocket states which is determined primarily by the
strength of the rotational state. From the Pauli principle, each rotational state of symmetry \textit{A,T} and \textit{E} is associated with a definite total spin of the four protons, $I=2,1,0 $ respectively. The energy levels of each state are 0, 0.16 and 0.24 \textit{meV}, respectively. The librational state starts at $E_{lib}=6.5 ~ meV$. Transitions between levels of different symmetry cannot be induced by phonon interaction alone, but are mediated by the weak dipole-dipole interaction among protons. As a consequence, the spin system shows slow thermal relaxation after a sudden change of the lattice temperature\cite{Friedrich:1996lr}.

\section{\label{sec:3} Neutron Cross Section model of Solid Methane}
A methane molecule in a crystal shows three different degrees of freedom in the solid phase. The rigid $H_{4}$ tetrahedron rotates about the central carbon atom. The whole molecule oscillates in the crystal lattice. Finally, the protons vibrate around their equilibrium positions.

The total neutron scattering cross section was evaluated from the scattering function
\begin{equation}\label{eq:1}
\frac{d^{2}\sigma}{d\Omega dE}=b^{2}\left(\frac{k_{f}}{k_{i}}\right)S(Q,\omega),
\end{equation}
where $b$ denotes the scattering amplitude and $k_{i}$ and $k_{f}$ the momentum of neutron before and after the scattering.

The scattering function $S(Q,\omega)$ contains the complete dynamics of the methane molecule;  rotation about the center, transition of the center of mass and molecular vibration. If there is no coupling among these modes, $S(Q,\omega)$  may be written as a convolution of the dynamic structure factors for  the different degrees of freedom of the motion:
\begin{equation}\label{eq:2}
S(Q,\omega)=S_{rot}(Q,\omega)\otimes S_{trans}(Q,\omega)\otimes S_{vib}(Q,\omega).
\end{equation}
By integrating $S(Q,\omega)$ over the solid angle $\Omega$  and the final neutron energy $E_{f}$, the total cross section $\sigma_{tot}$ is obtained.

If each degree of freedom is active  in different neutron energies, $S(Q,\omega)$  in Eq. \ref{eq:2} can be further approximated by three different terms depending on the incident neutron energy $E_{n}$.

For $E_{n}\leq 6.5 meV$, the protons in $CH_{4}$ behave like bound particles. Only the lowest few  rotational energy levels are thermally populated, while the molecules are in their translational and vibrational ground states \cite{grieger:3161}. Therefore, for the translational and vibrational motions, $S(Q,\omega)$  can be approximated with Debye-Waller forms   $\textrm{exp}(-\gamma_{trans} Q^2)$ and $\textrm{exp}(-\gamma_{vib} Q^2)$ respectively. The $\gamma$ stands for a model parameter $\gamma= \gamma_{rot} +\gamma_{trans} +\gamma_{vib}\simeq2.15\times10^{-4}nm^{2}$\cite{grieger:3161}.

Then the scattering function $S(Q,\omega)$ in this energy region can be approximated as
\begin{equation}\label{eq:3}
S(Q,\omega)=S_{rot}(Q,\omega)\textrm{exp}\left(-(\gamma_{trans}+\gamma_{vib})Q^2\right).
\end{equation}

For $6.5~ meV\leq E_{n}\leq 165~ meV$,  the $H_{4}$ tetrahedron reaches its rotational ``free" state and the librational degree of freedom is active.  The cross section for  rotations decreases, and the cross section for molecule oscillations in the crystal lattice appears ,
\begin{equation}\label{eq:4}
S(Q,\omega)=\textrm{exp}\left(-\gamma_{vib}Q^2\right)S_{rot}(Q,\omega)\otimes S_{trans}(Q,\omega).
\end{equation}

When neutrons reach the  energy $E_{n}=165 meV$, the protons start to behave like individual free particles. The neutron scattering is then dominated not only by multiple inelastic scattering, but also by vibration of the tetrahedrons. The rotational degree of freedom is now treated as a simple Debye-Waller form ,
\begin{equation}\label{eq:5}
S(Q,\omega)=\textrm{exp}\left(-\gamma_{rot}Q^2\right)S_{trans}(Q,\omega)\otimes S_{vib}(Q,\omega).
\end{equation}

\subsection{\label{sec:3-1} Rotational Excitation}
The rotational excitation of solid methane in phase II was taken  from Ozaki's work \cite{ozaki:3442}\cite{ozaki2} which has a more complete description of rotational excitation of methane. The double differential cross section for rotational motion can be divided into two parts:
\begin{equation}\label{eq:6}
\frac{d^{2}\sigma}{d\Omega dE}=\left(\frac{d^{2}\sigma}{d\Omega dE}\right)_{spin-ind}+~\left(\frac{d^{2}\sigma}{d\Omega dE}\right)_{spin-dep}.
\end{equation}
Each term of Eq. \ref{eq:6} represents coherent and incoherent  scattering in the rotational system as
\begin{equation}\label{eq:7}
\begin{split}
\left(\frac{d^{2}\sigma}{d\Omega dE}\right)_{spin-ind}=&~b_{coh}^{2}\frac{k_{f}}{k_{i}} p_{i}(T) g_{i}g_{f}\sum_{l=0}^{\infty}(2l+1)j_{l}^{2}(\textbf{k}\textbf{r})
\times F_{l}^{i, j}\delta(\hbar \omega+E_{f}-E_{i})\\
\left(\frac{d^{2}\sigma}{d\Omega dE}\right)_{spin-dep}=&~b_{inc}^{2}\frac{k_{f}}{k_{i}}p_{i}(T) g_{i}g_{f}\sum_{l=0}^{\infty}(2l+1)j_{l}^{2}(\textbf{k}\textbf{r})
\times G_{l}^{i, j}\delta(\hbar \omega+E_{f}-E_{i}),
\end{split}
\end{equation}
in which the coherent scattering corresponds to spin-independent part and incoherent part to spin-dependent one. $p_{i}(T)$ are the occupation probabilities for the  initial states in thermal equilibrium
\begin{equation}
p_{i}(T)=\frac{\mathrm{exp}(-E_{i}/k_{B}T)}{\displaystyle\sum_{i} \mathrm{exp}(-E_{i}/k_{B}T)}.\nonumber
\end{equation}

From Eq. \ref{eq:7}, the scattering functions of molecular rotation in each rotational state are
\begin{equation}\label{eq:23}
\begin{split}
S_{rot}^{coh}(Q,\omega)_{i,f}&=p_{i}(T) g_{i}g_{f} \sum_{l=0}^{\infty}(2l+1)j_{l}^{2}(\textbf{k}\textbf{r})\times F_{l}^{i, j}\delta(\hbar \omega+E_{f}-E_{i})\\
S_{rot}^{inc}(Q,\omega)_{i,f}&=p_{i}(T) g_{i}g_{f}\sum_{l=0}^{\infty}(2l+1)j_{l}^{2}(\textbf{k}\textbf{r})\times G_{l}^{i, j}\delta(\hbar \omega+E_{f}-E_{i}).
\end{split}
\end{equation}
Here $g_{i}$ and $g_{f}$ denote the degeneracy of the initial and final states and  $F_{l}^{i, j}$ and $G_{l}^{i, j}$ are transition integral coefficient between the initial and final states which were calculated in  Ozaki's work \cite{ozaki:3442}\cite{ozaki2}. $j_{l}(\textbf{k}\textbf{r})$ is the spherical Bessel function. The $\mathbf{r}$ stands for $C-H$ distance within a methane molecule ($\mathbf{r}=1.093\textrm{\AA}$).

After summing  over all initial and final states, the complete expression for the  scattering function for molecular rotations in  solid methane including coherent and incoherent parts is
\begin{equation}\label{eq:23}
S_{rot}(Q, \omega)=\sum_{i,f}(S_{rot}^{coh}(Q,\omega)_{i,f}+S_{rot}^{inc}(Q,\omega)_{i,f}).
\end{equation}

\subsection{\label{sec:level1} Intermolecular Vibration : Multi-phonon Excitation}
The incoherent scattering from a Bravais lattice due to one-phonon processes is
\begin{equation}\label{eq:8}
\frac{d^{2}\sigma}{d\Omega dE}=b_{inc}^{2} \sum_{q,s} \frac{k_{f}}{k_{i}}\delta(\hbar\omega\pm\hbar f) \left| \overline{\textbf{Q}}\cdot \overline{\textbf{V}}\right|^{2}
\times\frac{\hbar(n+\frac{1}{2}\pm\frac{1}{2})}{2Mf}\mathrm{e}^{-2W},
\end{equation}
where $\mathbf{Q}$ denotes momentum transfer and $\mathbf{V}$ is  the polarization vector for each phonon mode\cite{egelstaff-1}.

If $\left| \overline{\textbf{Q}}\cdot \overline{\textbf{V}}\right|^{2}$ in Eq.\ref{eq:8} is replaced by $\left| \overline{\textbf{Q}\cdot \textbf{V}}\right|^{2}$ where the bar denotes that an average over a constant frequency surface has been taken, the incoherent cross-section may be written as
\begin{equation}\label{eq:9}
\frac{d^{2}\sigma}{d\Omega dE}=b_{inc}^{2} \int_{0}^{\infty}df z(f)\frac{k_{f}}{k_{i}} \left| \overline{\textbf{Q}\cdot \textbf{V}}\right|^{2}\frac{\hbar(n+\frac{1}{2}\pm\frac{1}{2})}{2Mf}\mathrm{e}^{-2W},
\end{equation}
where $z(f_{i})$
is a frequency of \textit{i}th phonon and cubic symmetry has been assumed.

Since phonons of different wave vectors are independent in the incoherent approximation, the generalization of one phonon scattering to the multi-phonon case is straightforward.  We considered a number of different phonon modes changing their population by $\pm 1$ and thus n-phonon cross section contains a production of n-factors of type $\left| \overline{\textbf{Q}\cdot \textbf{V}}\right|^{2} \left| \phi \right|^{2}$.  It can thus be shown that the cross-section for  neutron downscattering is
\begin{equation}\label{eq:10}
\frac{d^{2}\sigma}{d\Omega dE}=b_{inc}^{2} N \frac{k_{f}}{k_{i}} \frac{1}{n!}\prod_{i=1}^{n} \left(\int_{-\infty}^{\infty}df \frac{z(f_{i})}{f_{i}} \frac{ \hbar^{2}Q^{2}/2M}{\mathrm{e}^{\frac{\hbar f_{i}}{k_{i}T}-1}}\right)
\times\mathrm{e}^{-2W}\delta(\hbar\omega+\sum_{i}\hbar f_{i}).
\end{equation}
Using the Debye approximation, it is possible to carry out the necessary integrations to obtain the cross-section $\sigma_{n}^{inc}$. The total inelastic incoherent cross-section is then given by
\begin{equation}\label{eq:11}
\sigma_{inelastic}^{inc}=\sum_{n=1}^{\infty}\sigma_{n}^{inc}.
\end{equation}

The scattering function $S_{trans}(Q,\omega)$ is now written as
\begin{equation}\label{eq:12}
\begin{split}
S_{trans}(Q,\omega)&=\sum_{n=1}^{\infty}S^{n}_{trans}(Q,\omega)\\
&=\sum_{n=1}^{\infty}\frac{1}{n!}\prod_{i=1}^{n} \left(\int_{-\infty}^{\infty}df \frac{z(f_{i})}{f_{i}} \frac{ \hbar^{2}Q^{2}/2M}{\mathrm{e}^{\frac{\hbar f_{i}}{k_{i}T}-1}}\right)\quad \times\mathrm{e}^{-2W}\delta(\hbar\omega+\sum_{i}\hbar f_{i})
\end{split}
\end{equation}

\subsection{\label{sec:3-2} Intramolecular Vibrations}
From the simple harmonic oscillator model\cite{Lovesey}, the scattering function for intramolecular vibration is
\begin{equation}\label{eq:13}
S(Q, \omega)=\frac{1}{2\pi\hbar}\int_{-\infty}^{\infty}dt ~\mathrm{exp}(-i\omega t)
\times \left< \mathrm{exp}(-i \mathbf{Q}\cdot\hat{\mathbf{r}})\mathrm{exp}\left\{i \mathbf{Q}\cdot\hat{\mathbf{r}}(t)\right\}\right>.
\end{equation}

To calculate the response function for our  model, we need to calculate the correlation function,
\begin{equation}\label{eq:104}
\begin{split}
\left< \mathrm{exp}(-i \mathbf{Q}\cdot\hat{\mathbf{r}})\mathrm{exp}\left\{i \mathbf{Q}\cdot\hat{\mathbf{r}}(t)\right\}\right>&=\prod_{\alpha=x,y,z}\left< \mathrm{exp}(-i \mathbf{Q}_{\alpha}\cdot\hat{\mathbf{r}}_{\alpha})\mathrm{exp}\left\{i \mathbf{Q}_{\alpha}\cdot\hat{\mathbf{r}}_{\alpha}(t)\right\}\right> \\
&=\mathrm{exp}\left\{-\left<(\mathbf{Q}\cdot\hat{\mathbf{r}})\right>+\left<(\mathbf{Q}\cdot\hat{\mathbf{r}})\left\{\mathbf{Q}\cdot\hat{\mathbf{r}}(t)\right\}\right>\right\}.
\end{split}
\end{equation}

For an isotropic harmonic oscillator,
\begin{equation}\label{eq:15}
\left<(\mathbf{Q}\cdot\hat{\mathbf{r}})\left\{\mathbf{Q}\cdot\hat{\mathbf{r}}(t)\right\}\right>=\frac{\hbar Q^{2} \mathrm{cosh}\left\{\omega_{0}(it+\frac{1}{2}\hbar\beta)\right\}}{2M\omega_{0} \mathrm{sinh}\left(\frac{1}{2}\hbar\omega_{0}\beta\right)}.
\end{equation}

On setting $t=0$, we find on the right side of the Eq. \ref{eq:15},
\begin{equation}\label{eq:16}
\left< ~\left(\mathbf{Q}\cdot\hat{\mathbf{r}}\right)^{2}\right>=\left(\frac{\hbar Q^{2}}{2M \omega_{0}}\right)\mathrm{coth}\left(\frac{1}{2}\hbar \omega_{0}\beta\right)=2W(Q).
\end{equation}

Using the identity,
\begin{equation}\label{eq:17}
\mathrm{exp}(y)\mathrm{cosh}(x)=\sum_{-\infty}^{\infty}\mathrm{exp}(nx)I_{n}(y)
\end{equation}
where $n$ is an integer and $I_{n}(y)=I_{n}(-y)$ is a Bessel function of the first kind. Using $x=\omega_{0}(it+\frac{1}{2}\hbar\beta)$ and $y=\frac{\hbar Q^{2}}{2M \omega_{0} \mathrm{sinh}\left(\frac{1}{2}\hbar \omega\beta\right)}$, the correlation function becomes
\begin{equation}\label{eq:18}
\mathrm{exp}\left< \left(\mathbf{Q}\cdot\hat{\mathbf{r}}\right)\left\{\mathbf{Q}\cdot\hat{\mathbf{r}}(t)\right\}\right>=\sum_{-\infty}^{\infty}I_{n}(y)\mathrm{exp}\left\{n \omega_{0}(it+\frac{1}{2}\hbar\beta)\right\}.
\end{equation}
and  we obtain the scattering function as
\begin{equation}\label{eq:19}
S(Q,\omega)=\mathrm{exp}(-2W(Q)+\frac{1}{2}\hbar\omega\beta)\sum_{-\infty}^{\infty}I_{n}(y)\delta(\hbar\omega-n\hbar\omega_{0})
\end{equation}
where $\omega_{0}$ is normal vibration frequency. A methane molecule has four normal vibrations with characteristic energies of $\hbar\omega_{0}=168, 172, 376$ and $391 meV$ \cite{1951PhRv...84..204M}.

\begin{figure}[htbp]
\centering
\subfigure[20K] 
{
    \label{fig:2:a}
    \includegraphics[width=10cm]{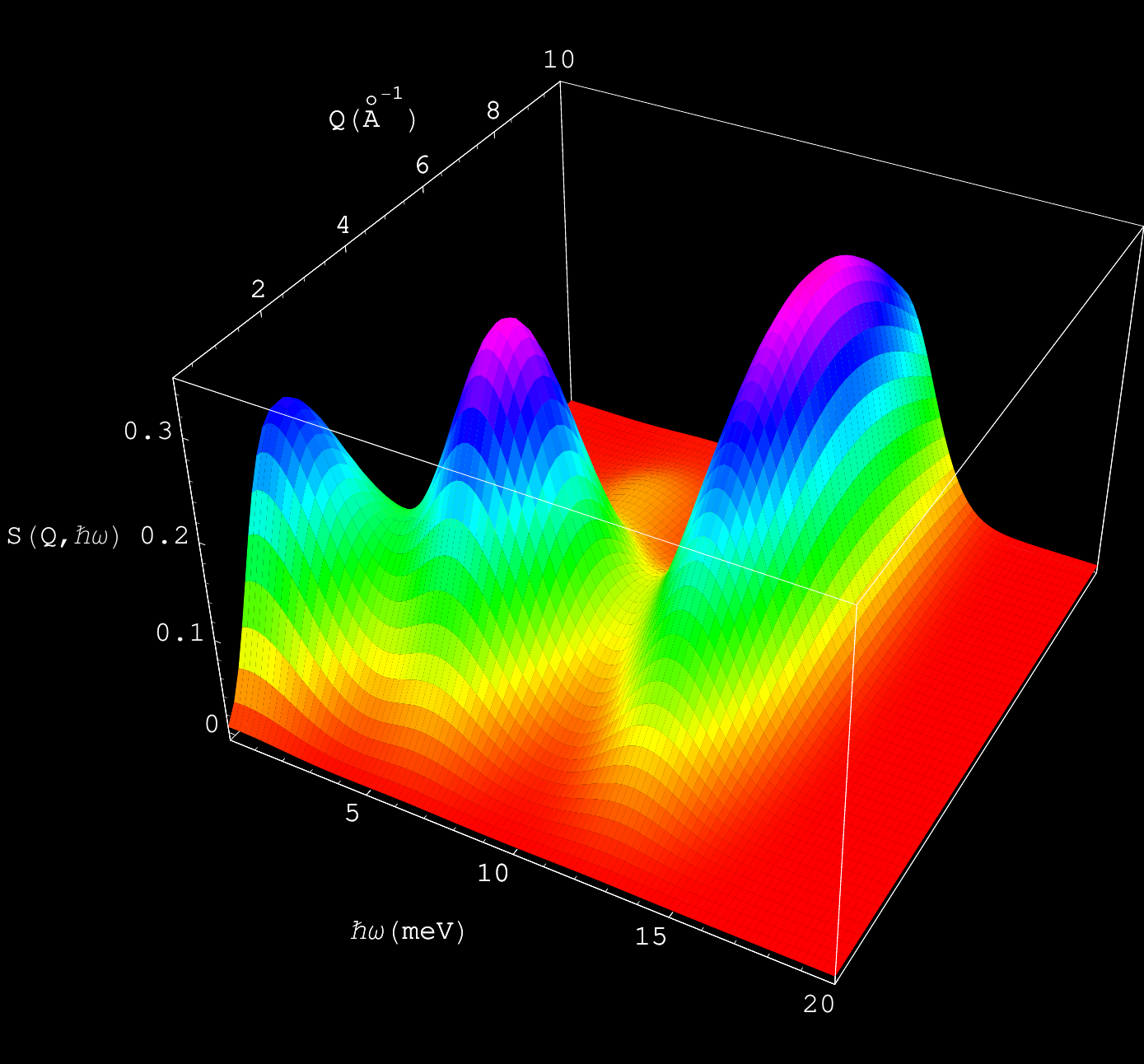}
}
\hspace{1cm}
\subfigure[4K] 
{
    \label{fig:2:b}
    \includegraphics[width=10cm]{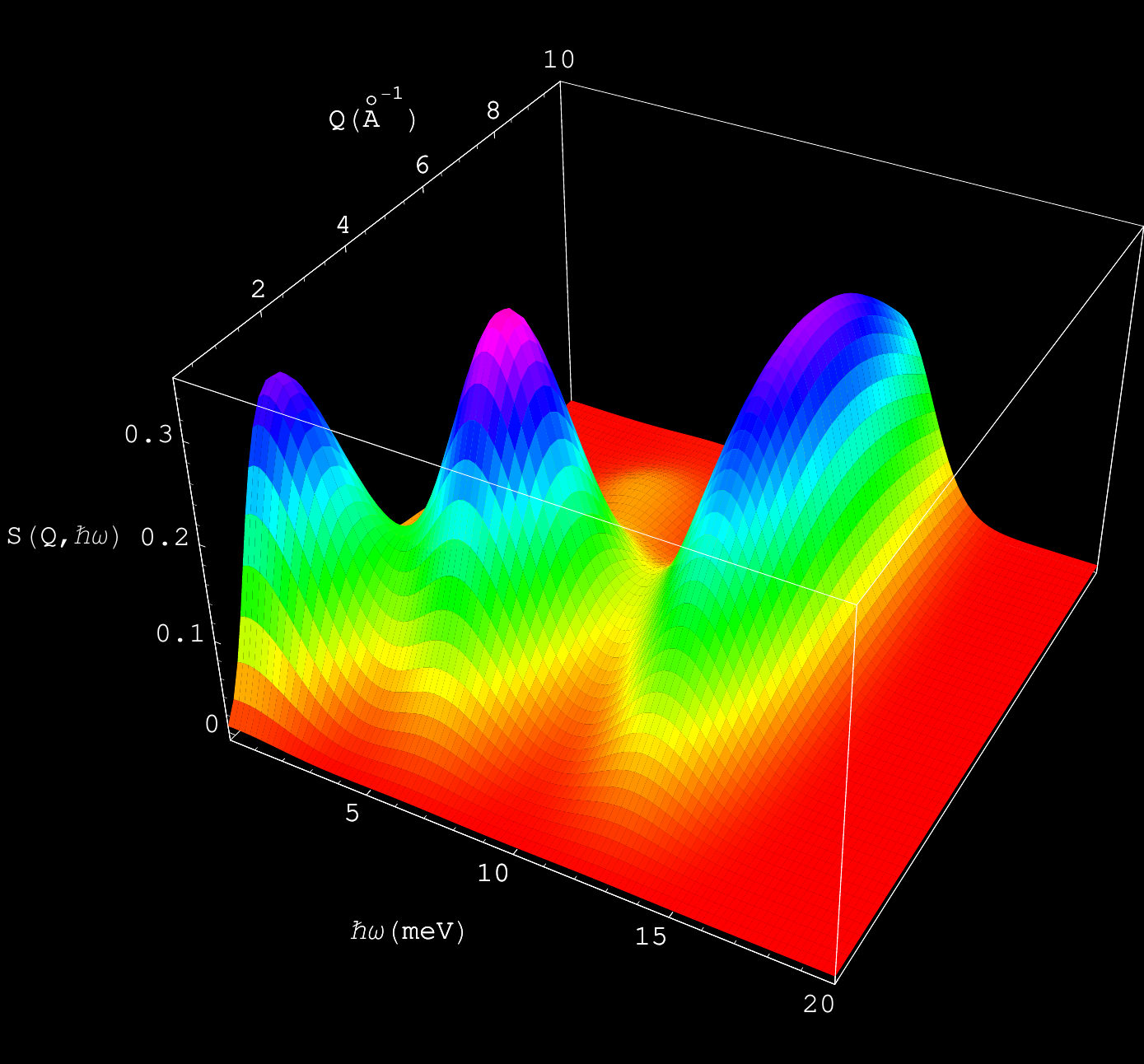}
    }
\caption{The 3D plots of scattering function $S(Q,\omega)$ in 20K and 4K.}
\label{fig:2} 
\end{figure}

\section{\label{sec:4} Result and Discussion}

\subsection{\label{sec:4-1} Scattering Function $S(Q, \omega)$ of Solid Methane}

The total scattering function is the convolution of scattering functions of the  three different modes. As mentioned, we treat this total scattering function as the linear combination of each scattering function in Eq. \ref{eq:1}, Eq. \ref{eq:2} and Eq. \ref{eq:3} depending on incident neutron energy. This is an  approximation to calculate the total scattering function when  each different scattering mode dominates in different neutron energy regimes. We show the 3D plot of $S(Q, \omega)$ in 20K and 4K  as a function of  momentum transfer ($Q=0\sim 10\textrm{\AA}^{-1}$) and energy transfer ($\hbar\omega=0\sim 20 meV$) in Fig. \ref{fig:2:a} and Fig. \ref{fig:2:b}. In these figures, the two peaks in lower energy transfer ($ 0 \lesssim \hbar\omega \lesssim10 meV$) represent the densities of energy and momentum transfers from the rotational modes. The long knoll around $\hbar\omega \sim 12meV$ denotes the density due to the multi-phonon excitation . For  lower temperature, the overall density for  energy transfer from the  rotational mode is increased. The energy transfer to  phonons  is essentially unchanged.

\subsection{\label{sec:4-2} Generalized Frequency Spectrum}
It has been remarked that the important part of the scattering law for neutron thermalization problem is the self-term defined as Van-Hove self-correlation function
\begin{equation}\label{eq:20}
G_{s}(r,\tau)=\frac{1}{N}\left<{\sum_{n}\int_{v}dr^{'}\delta(r+r_{n}(0)-r^{'})\delta(r^{'}-r_{n}(\tau))}\right>
\end{equation}
and the scattering function have
\begin{equation}\label{eq:21}
S(Q,\omega)=\frac{1}{2\pi}\int \mathrm{e}^{i(\mathbf{Q}\cdot\mathbf{r}-\mathbf{\omega}\cdot\mathbf{\tau})}G_{s}(r,\tau)dr d\tau.
\end{equation}

Eq. \ref{eq:20} is the velocity correlation function for the motion of an atom in the medium. The physical definition of this function is as follows:
one images that the velocity of an atom is measured at a certain time $t_{1}$. Then the atom is followed as it moves about the system and its velocity is remeasured at a later time $t_{2}$. Finally, the product of these two velocity measurements is taken and averaged over all values of the velocity which may be generated by thermal excitation of the system\cite{egelstaff-2},
\begin{equation}\label{eq:22}
\begin{split}
\left(\frac{\omega^{2} \mathrm{exp}\left(-\frac{\hbar\omega}{2k_{B}T}\right)S(Q,\omega)}{Q^{2}}\right)_{Q\rightarrow0}
&=\frac{\mathrm{exp}\left(-\frac{\hbar\omega}{2k_{B}T}\right)}{2\pi} \int_{-\infty}^{\infty}\left<v_{Q}(0)v_{Q}(\tau)\right>\mathrm{e}^{i\omega\tau}d\tau\\
&=\frac{k_{B}T}{2M}P(\omega),
\end{split}
\end{equation}
which defines the function $P(\omega)$. The velocity correlation function is a complex function and its real and imaginary parts are related to the function $P(\omega)$ as
\begin{align}\label{eq:23}
\mathrm{Im}\left< v_{Q}(0)v_{Q}(\tau)\right>&=\frac{\hbar}{2M}\int_{0}^{\infty}\omega \rho(\omega) \mathrm{sin} (\omega\tau)d\omega\nonumber\\
\mathrm{Re}\left< v_{Q}(0)v_{Q}(\tau)\right>&=\frac{k_{B}T}{2M}\int_{0}^{\infty}\frac{P(\omega) \mathrm{cos} (\omega\tau)}{\mathrm{cosh}({\hbar\omega}/{2k_{B}T})}d\omega,
\end{align}
where $\rho(\omega) =P(\omega)\frac{\mathrm{sinh}({\hbar\omega}/{2k_{B}T})}{{\hbar\omega}/{2k_{B}T}}$ is the generalized frequency spectrum and it has a condition to be $\int_{0}^{\infty}\rho(\omega)d\omega=1$.
The function  $\rho(\omega)$ can be applied in many cases and  it is the same as  the frequency distribution in a solid. But in the case of any other material, it depends on the types of motion which are going on in the system. In solid methane, the frequency spectrum represents all characteristics of the rotational, translational and vibrational behaviors of a molecule\cite{egelstaff-3}.

\begin{figure}[htbp]
\begin{center}
\includegraphics[width=10cm]{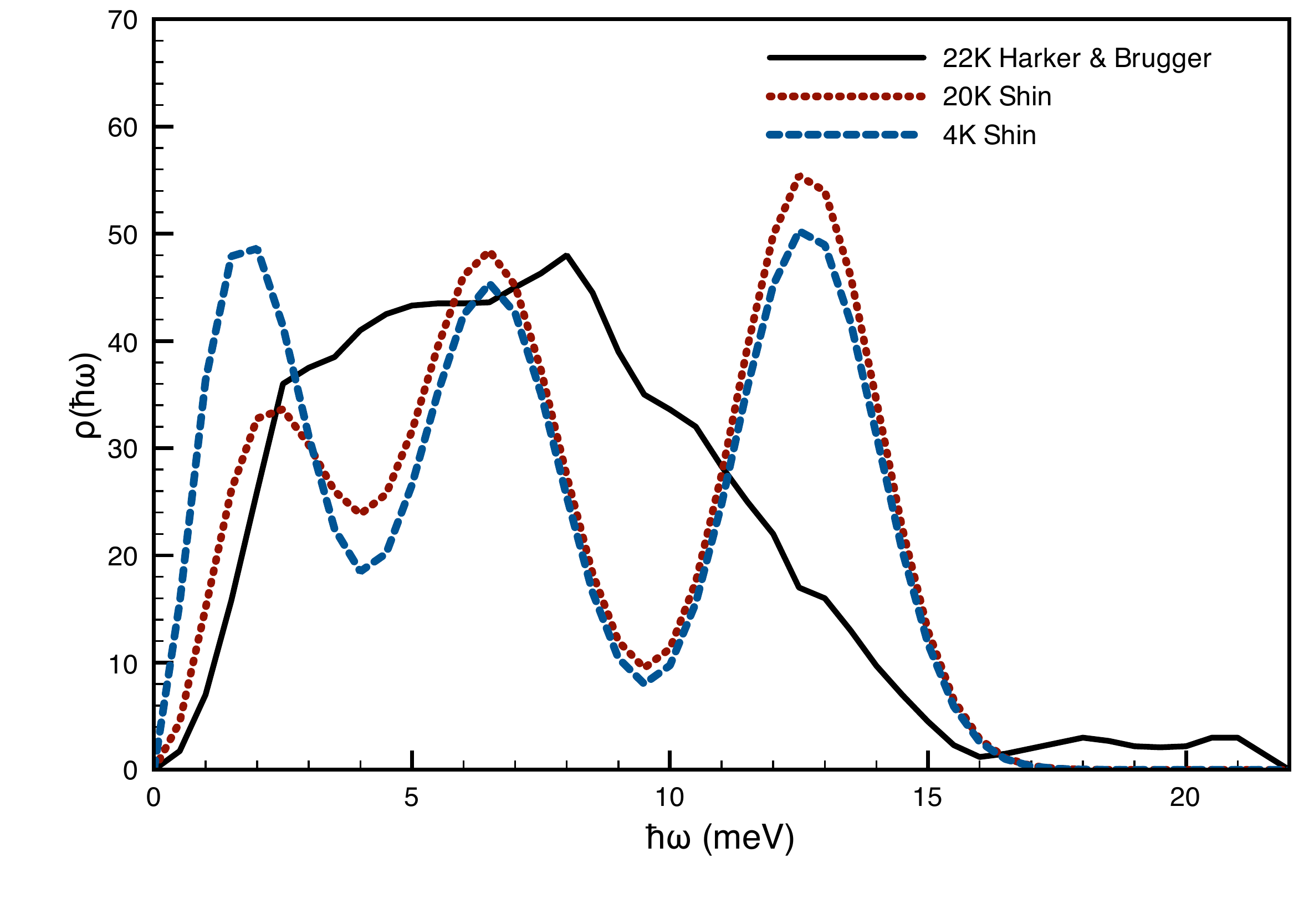}
\caption{The frequency spectrum of solid methane in the 20K and 4K temperatures. The Harker $\&$ Brugger spectrum was in the 22K}
\label{fig:1}
\end{center}
\end{figure}

Fig. \ref{fig:1} shows the frequency spectra of solid methane in two phase II temperatures, 20K and 4K, compared with the 22K Harker $\&$ Brugger's frequency spectrum\cite{HARKER1967}. The black solid line is the Harker $\&$ Brugger's frequency spectrum in 22K which was generated by  interpolation  from  data. The red dotted line is the frequency spectrum at  20K and the blue dashed line is at  4K.  The first peak around 2 \textit{meV} in the frequency spectrum in 4K and 20K represents the linear superposition of the energy transitions from ground and the first excited rotational energy states to higher states in free rotational mode and the tunneling among ground state multiplets (three lowest states) in hindered rotational mode. The second peak denotes the linear superposition of the energy transition to the higher rotational states in free rotational mode and the librational states in the hindered rotational mode. The final peak around 12.9 \textit{meV} is a  multi-phonon mode  corresponding to the  Debye energy ($k_{B} T_{D}$). The multi-phonon scattering and higher excitations are suppressed at  low temperature, but the  transitions probabilities from the  free rotor state and ground state multiplets in hindered rotor increase.

\subsection{\label{sec:4-3} Total Neutron Scattering Cross Section}
The total neutron scattering cross sections have not only been calculated from the double-differential cross section in Eq. \ref{eq:1},  but also generated from the frequency spectra employing LEAPR module of NJOY code.

In the low neutron energy limit, the scattering is mainly elastic scattering from free and hindered rotors and from tunneling among multiplet of rotational ground states in hindered rotor. We include the elastic coherent and incoherent scattering of hydrogen  from Ozaki \cite{ozaki:3442}\cite{ozaki2} and Grieger\cite{grieger:3161}.

Around $E_{n}=10 meV$, the total cross-section approaches  200$b$ in two different temperatures. For $E_{n}\leq 10 meV$, $\sigma_{tot}$ mainly depends on translational excitation. In the higher energy region, the vibrational excitation becomes more important than the rotational and translational scattering.

\begin{figure}[htbp]
\centering
\subfigure[20K] 
{
    \label{fig:3:a}
    \includegraphics[width=10cm]{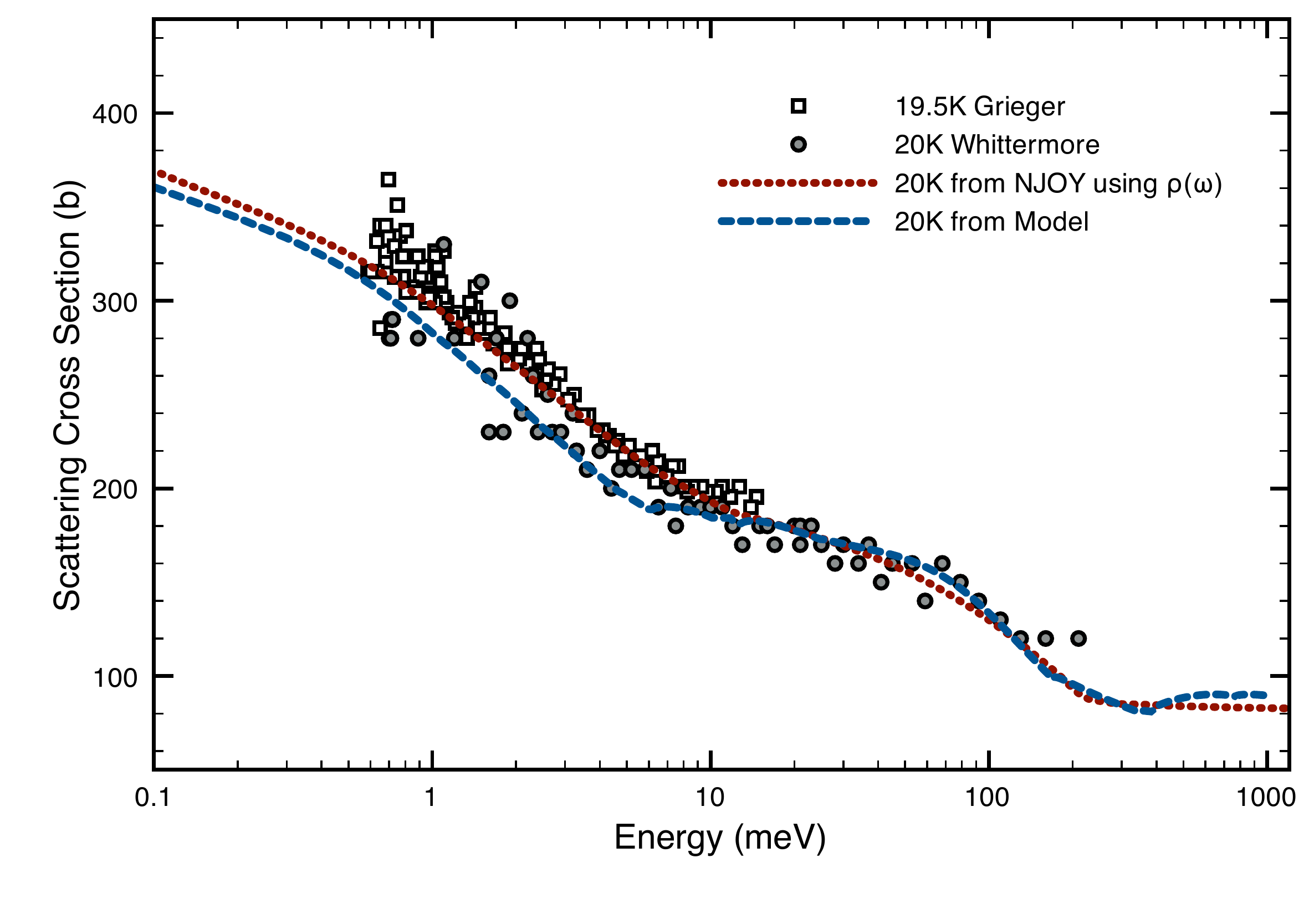}
}
\hspace{1cm}
\subfigure[4K] 
{
    \label{fig:3:b}
    \includegraphics[width=10cm]{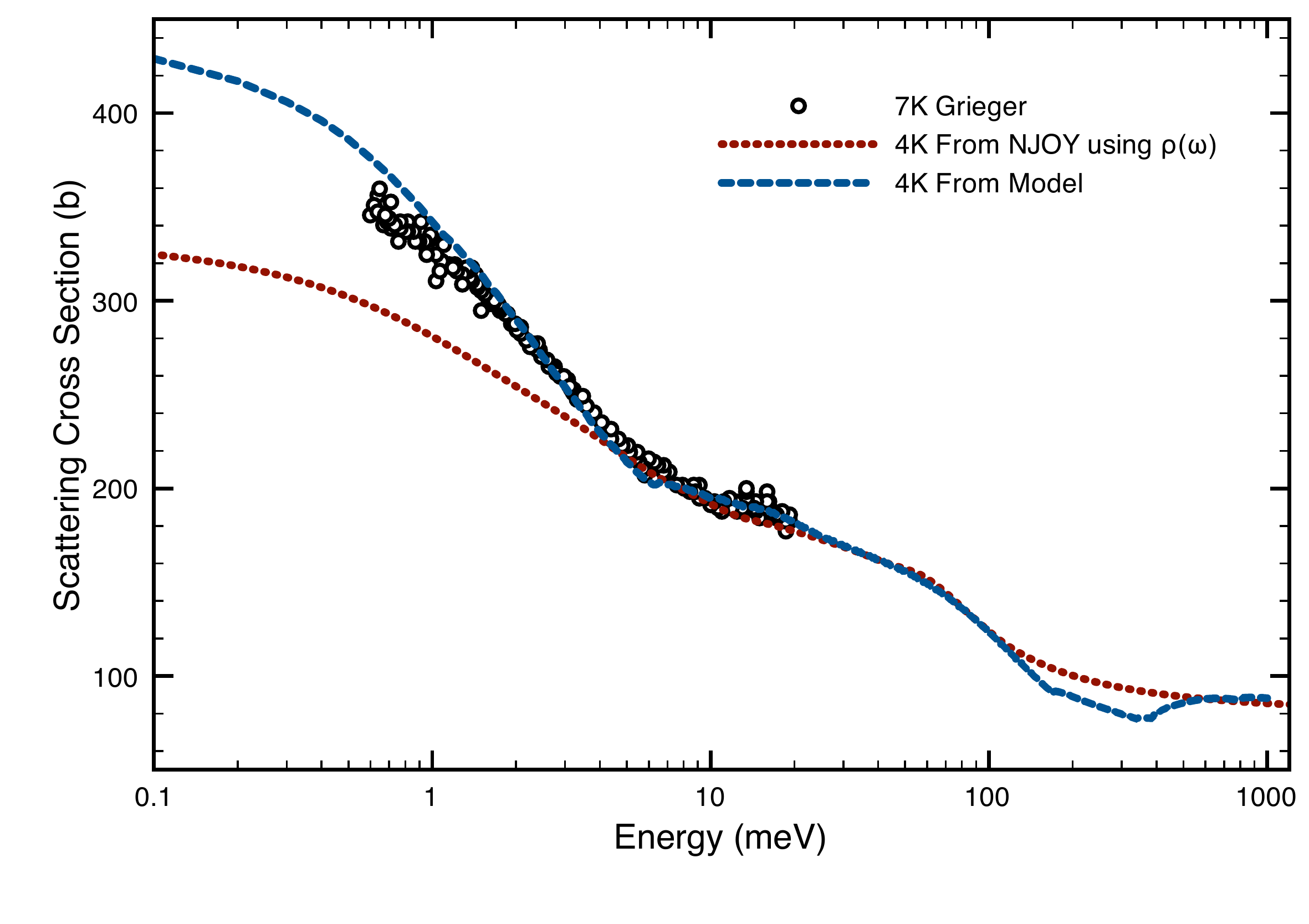}
}
\caption{The total neutron scattering cross section of solid methane in 20K and 4K. The NJOY curve  is from the cross-section generated from the LEAPR module in NJOY through the multi-phonon expansion method }
\label{fig:3} 
\end{figure}

Fig. \ref{fig:3:a} and \ref{fig:3:b} show the total neutron scattering cross section in 20K and 4K. In 20K, the theoretical calculation and the cross section generated from the frequency spectra through LEAPR module are compared to 20K measurement data \cite{Wittemore}. The frequency spectrum was used directly in multi-phonon expansion in the LEAPR module.
\begin{figure}[htbp]
\begin{center}
\includegraphics[width=10cm]{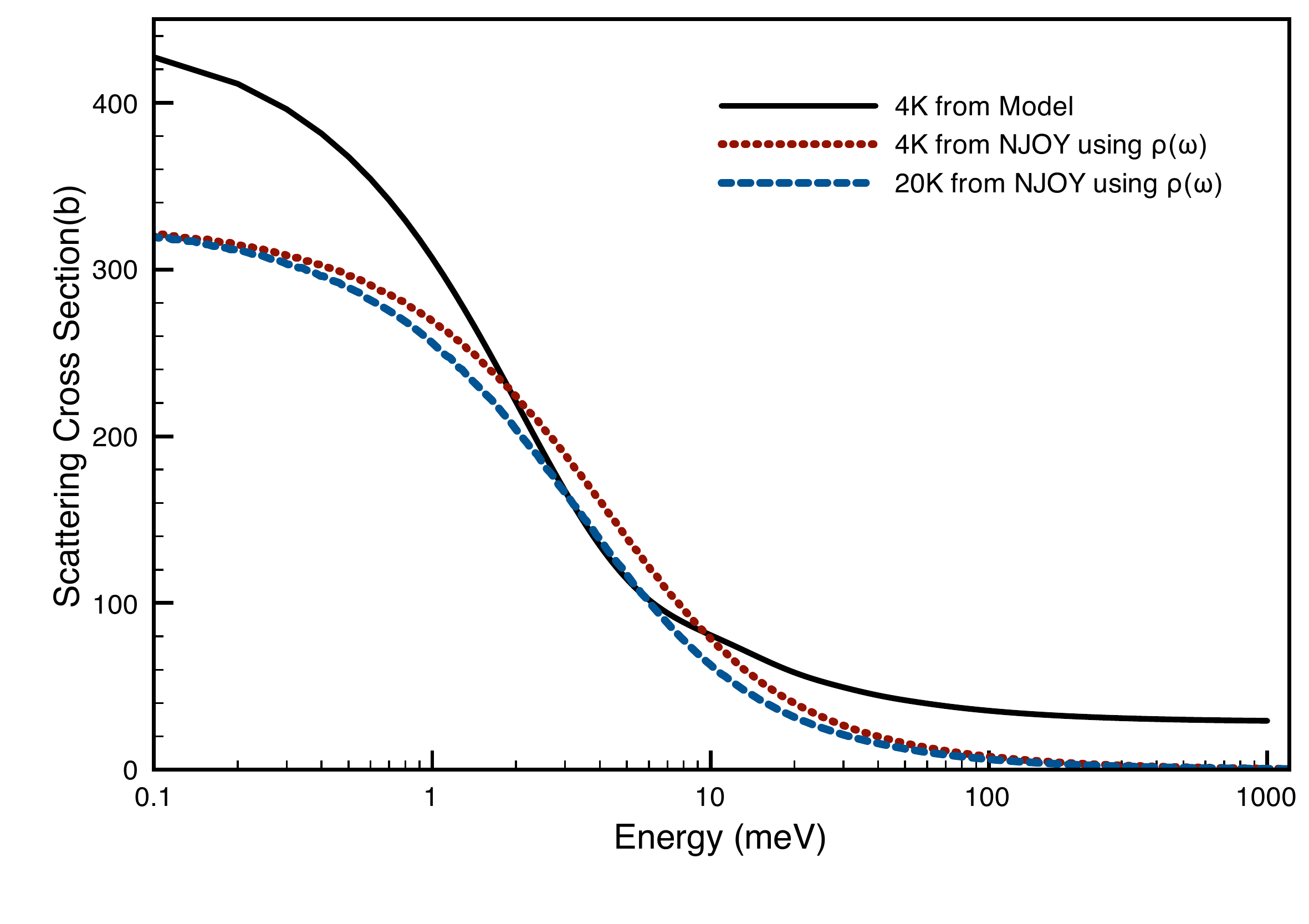}
\caption{The Elastic Scattering Cross Section in 4K and 20K from the LEAPR module and 4K from the model calculation.}
\label{fig:5}
\end{center}
\end{figure}
In Fig. \ref{fig:3:a}, the  black solid line is the cross section calculated from the model and the red dotted  line and blue dashed line are  the cross-sections at  4K and 20K which NJOY generated from the frequency spectrum. Although there is a certain difference between these two cross-sections, the overall cross-section of model and NJOY agree well with experimental data in 20K. In the 4K cross-section plot, however, there is a difference between theoretical model calculation and NJOY one. This is primarily due to the difference of elastic scattering cross-section. From the measurements and the theoretical calculation, it was clear that as the temperature is lowered, the elastic cross-section $\sigma_{el}$ should be increased and inelastic cross-section $\sigma_{inel}$ is decreased. However, the elastic scattering cross-section generated from NJOY wasn't changed even in the lower temperature which causes about 100$b$ of discrepancy between the theoretical calculation and NJOY across 0.1$meV$ neutron energy in the 4K.

\subsection{\label{sec:4-4} Decoupling of the Rotational Modes}
Despite of many speculations, as we mentioned in Chapter 1,  there has been no clear explanation on which one, between free and hindered rotations, is more effective in improving the cold neutron neutron flux. To approach to the answer, we assume that the one of the rotational modes could be turned off so that solid methane molecule has either free or hindered mode only in its rotational degree of freedom in our model. In addition, the multi-phonon scattering and vibrational mode are still in present. From the assumption, the scattering function, frequency spectrum and neutron scattering cross section in each case are compared .

\begin{figure}[htbp]
\centering
\subfigure[Free Rotationl and Phonon Modes] 
{
    \label{fig:4:a}
    \includegraphics[width=10cm]{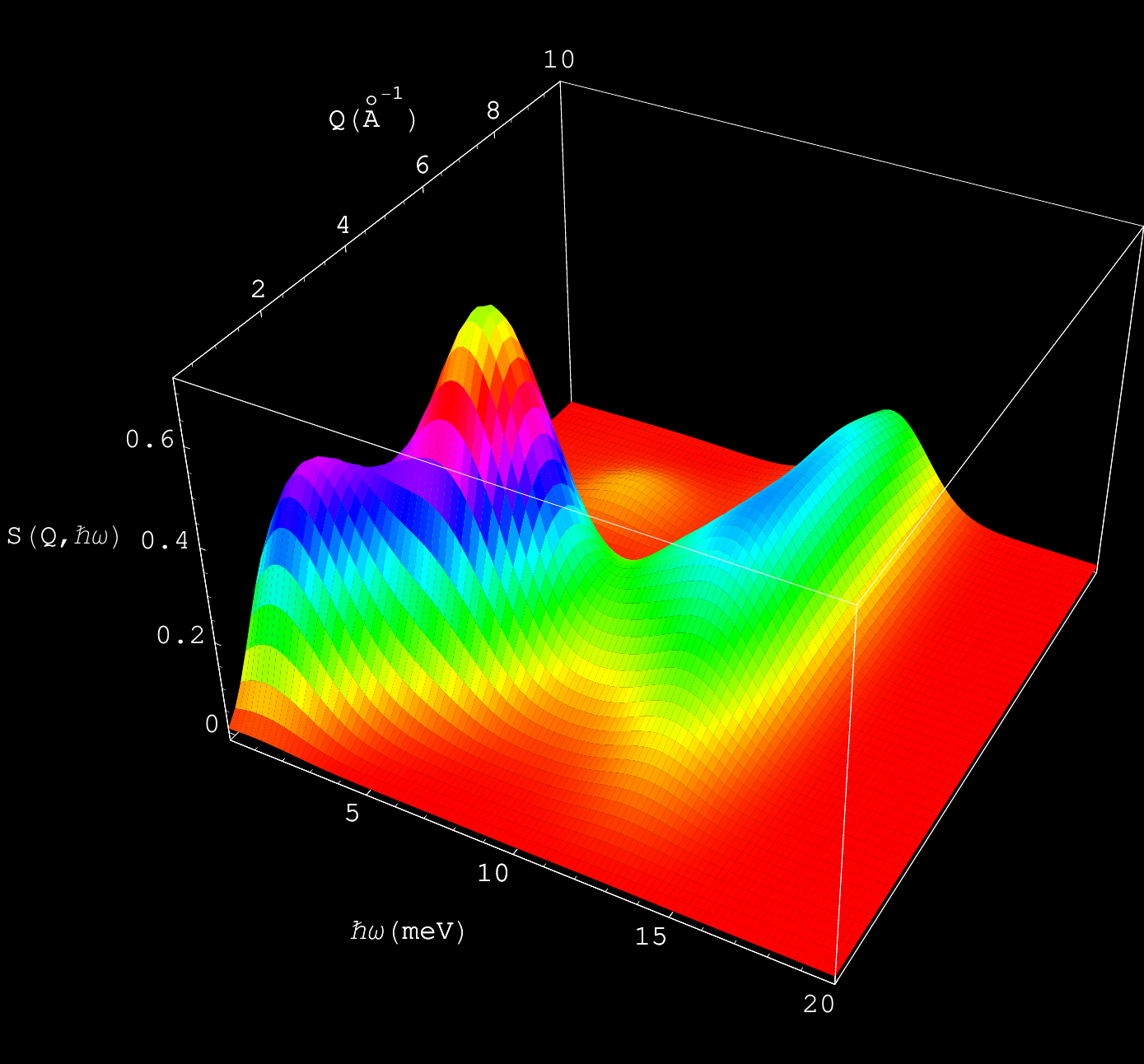}
}
\hspace{2cm}
\subfigure[Hindered Rotation and Phonon Modes] 
{
    \label{fig:4:b}
    \includegraphics[width=10cm]{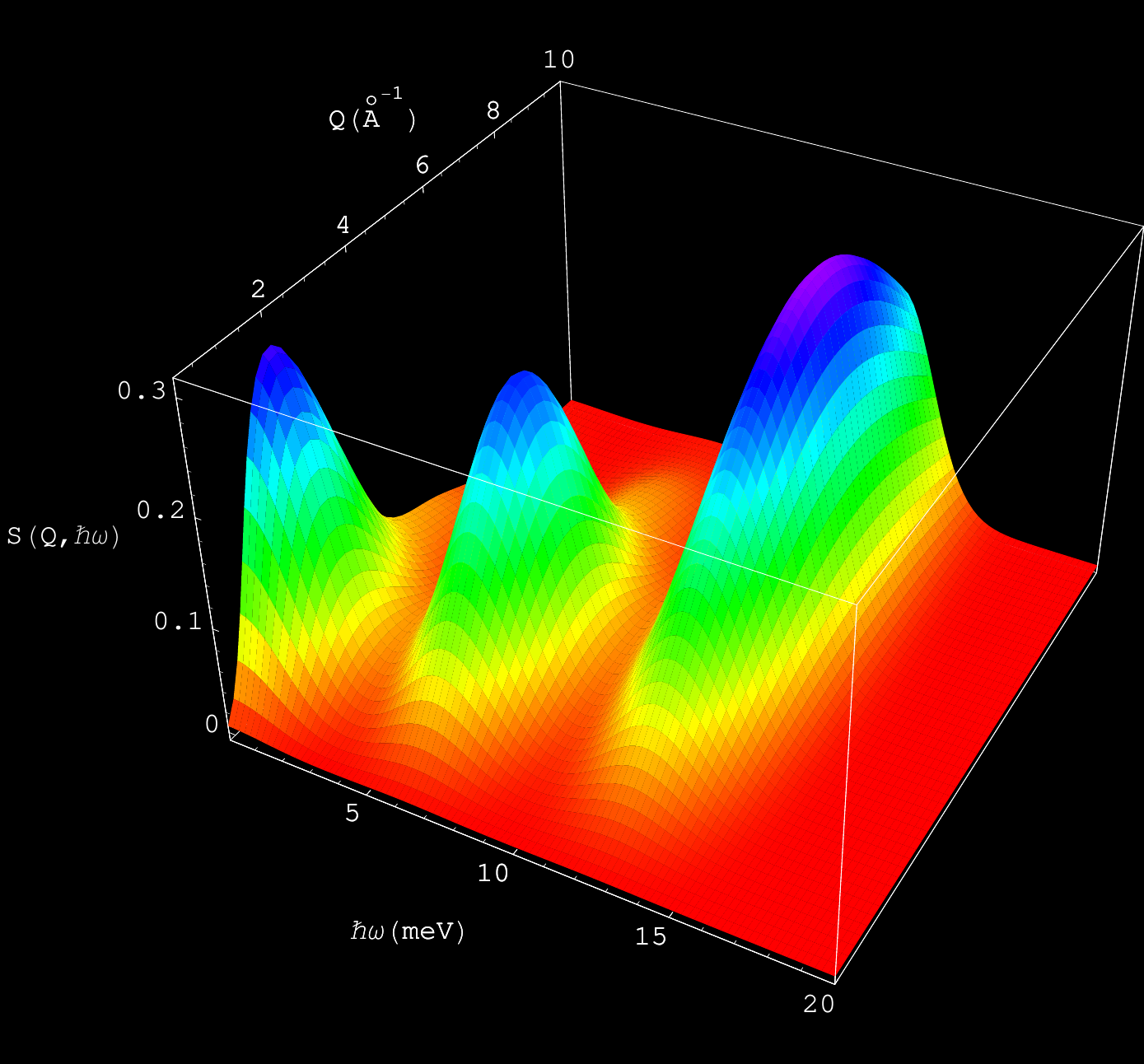}
}
\caption{The 3D plots of scattering function $S(Q,\omega)$ in 4K.}
\label{fig:4} 
\end{figure}

\begin{figure}[htbp]
\begin{center}
\includegraphics[width=10cm]{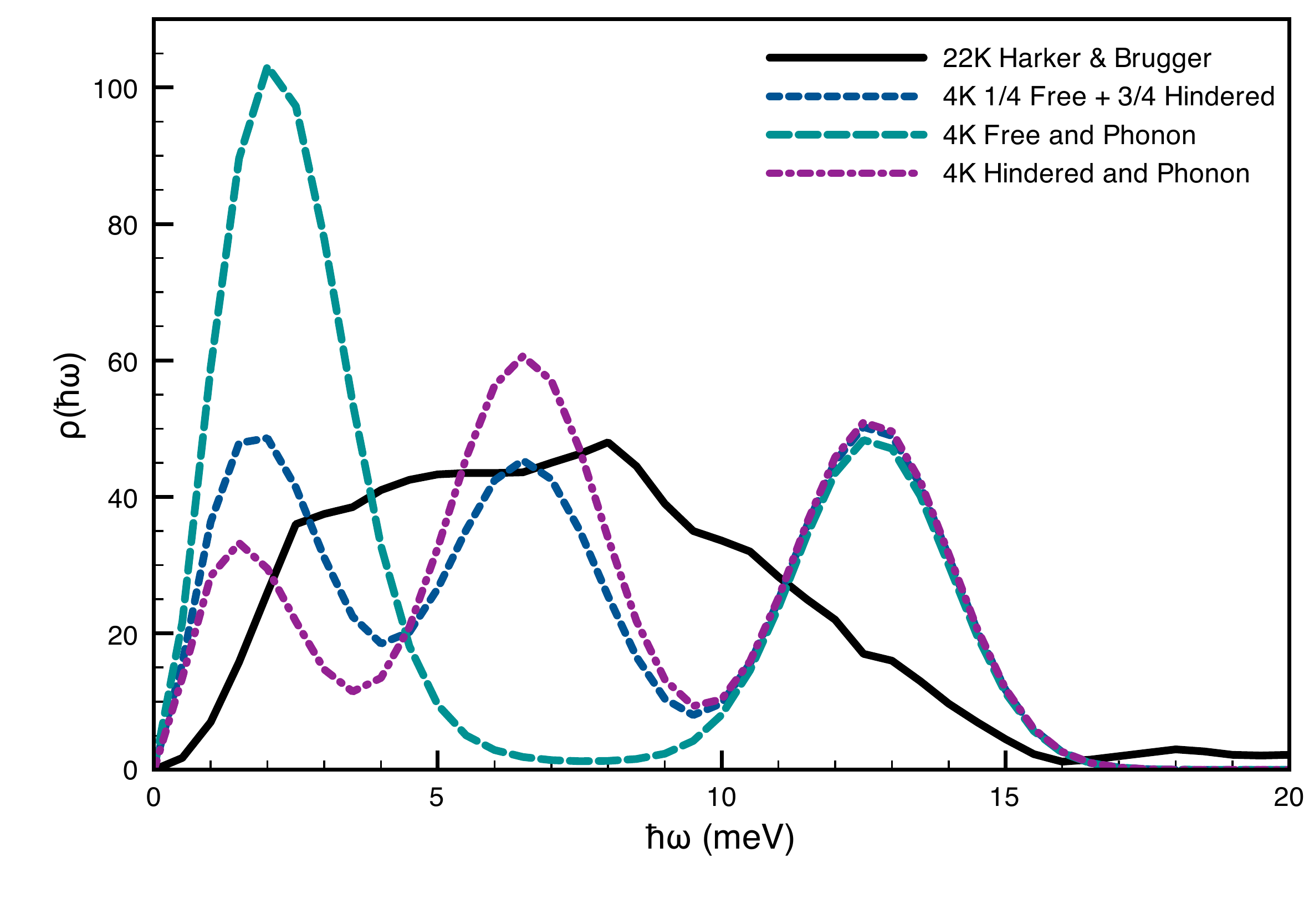}
\caption{The frequency spectrum of solid methane in the 4K temperatures. The Harker $\&$ Brugger spectrum was in the 22K}
\label{fig:5}
\end{center}
\end{figure}

\begin{figure}[htbp]
\begin{center}
\includegraphics[width=10cm]{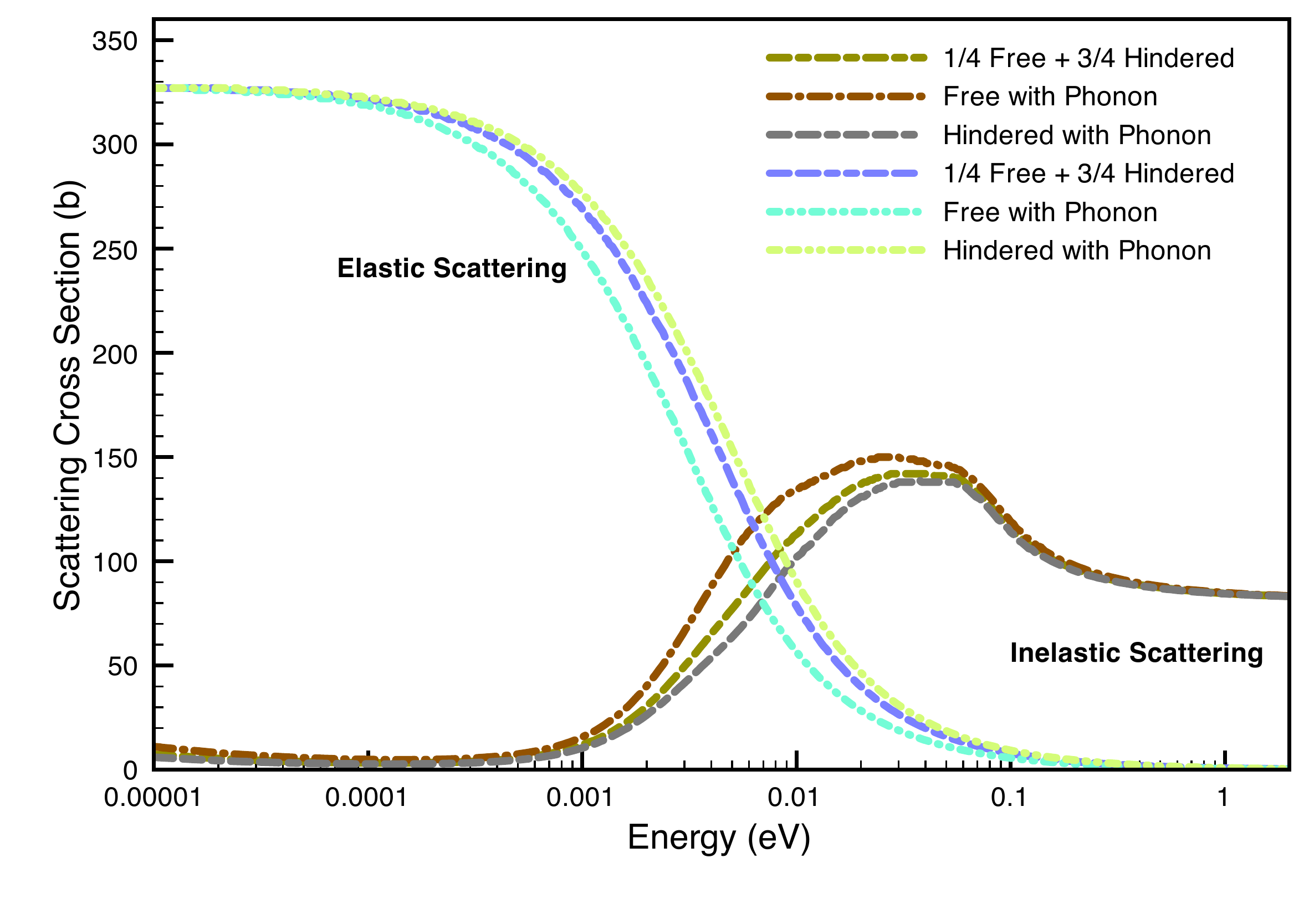}
\caption{The Scattering Cross Section of solid methane in the 4K temperatures.}
\label{fig:6}
\end{center}
\end{figure}

Fig. \ref{fig:4}, \ref{fig:5} and \ref{fig:6} show the scattering function, frequency spectrum and neutron scattering cross section of each cases in 4K system temperature. In the $S(Q,\omega)$ plots, the higher density of energy transfer are mainly concentrated in low energy area for free only rotational mode. The hindered rotation mode shows two separate distributions corresponding the rotational tunneling and the librations in high energy region. These features are consistent in the frequency spectrum plot. The free rotation mode has about twice higher density of low energy transfer than  hindered mode in $Q\rightarrow 0 $ limit. The density of librations in hindered rotations are also low. This higher density of inelastic scattering of free rotation made shows better neutron inelastic scattering cross section in Fig.\ref{fig:6}. Even though there has been no report to get all free up rotational modes of solid methane in phase II, it is clear from our studies that the free rotational mode of solid methane in phase II certainly show higher probability than hindered rotation mode in improving the cold neutron flux.

\section{\label{sec:level1} Conclusions}

Theoretical studies of the neutron scattering cross section model of solid methane were performed. Based on the model, scattering function $S(Q,\omega)$ and frequency spectrum $\rho(\omega)$ for solid methane in 20K and 4K are evaluated. The LEAPR module of NJOY code was also used for the evaluation of total scattering cross section from the frequency spectrum in the 20K and 4K. The total scattering cross sections are shown to be consistently explained in our model.  The role of each rotational mode was estimated in the scattering function, frequency spectrum and scattering cross section by switching off the other rotational mode. A neutron energy spectrum measurements and  the MCNP simulations will be made to validate our scattering model.

\section{\label{sec:level1} Acknowledgments}
This work was supported by the National Scientific Foundation under Grant No. DMR-0220560 and DMR-0320627.
\newpage
\bibliography{paper-yun}

\end{document}